\documentclass[preprint,twocolumn,showpacs,preprintnumbers,amsmath,amssymb]{revtex4}

\usepackage{graphicx}
\usepackage{dcolumn}
\usepackage{bm}
\usepackage{multirow}
\usepackage{xcolor}
\usepackage[colorlinks=true,linkcolor=black,citecolor=red,urlcolor=black]{hyperref}
\usepackage[sort&compress]{natbib}

\newcommand{\kCl}{$\kappa$-Cl}
\newcommand{\kBr}{$\kappa$-Br}

\begin{document}

\title{Low-frequency charge-carrier dynamics in the vicinity of the Mott transition in quasi-two-dimensional organic superconductors $\kappa$-(BEDT-TTF)$_2$X}
\author{Jens Brandenburg}\email{brandenburg@physik.uni-frankfurt.de}
\affiliation{Physikalisches Institut, Goethe-Universit\"at Frankfurt, 60438 Frankfurt(M), Germany}
\affiliation{Max Planck Institute for Chemical Physics of Solids, 01187 Dresden, Germany}
\author{Jens M\"uller}
\affiliation{Physikalisches Institut, Goethe-Universit\"at Frankfurt, 60438 Frankfurt(M), Germany}
\author{Dieter Schweitzer}
\affiliation{3.\ Physikalisches Institut, Universit\"at Stuttgart, 70569 Stuttgart, Germany}
\author{John A.\ Schlueter}
\affiliation{Argonne National Laboratory, Materials Science Division, Argonne, IL 60439, USA}

\date{\today}

\begin{abstract}
We investigate the dynamics of correlated charge carriers in the vicinity of the Mott metal-to-insulator transition in various of the title quasi-two-dimensional organic charge-transfer salts by means of fluctuation (noise) spectroscopy. The observed $1/f$-type fluctuations are quantitatively very well described by a phenomenological model based on the concept of non-exponetial kinetics. The main result is a correlation-induced enhancement of the fluctuations accompanied by a substantial shift of spectral weight to low frequencies close to the Mott critical endpoint. This sudden slowing down of the electron dynamics may be a universal feature of metal-to-insulator transitons. For the less correlated metallic materials, we find a crossover/transition from hopping transport of more-or-less localized carriers at elevated temperatures to a low-temperature regime, where a metallic coupling of the layers allows for coherent interlayer transport of delocalized electrons. 
\end{abstract}


\maketitle

\begin{figure}[htbp]
\begin{center}
\includegraphics[width=0.475\textwidth]{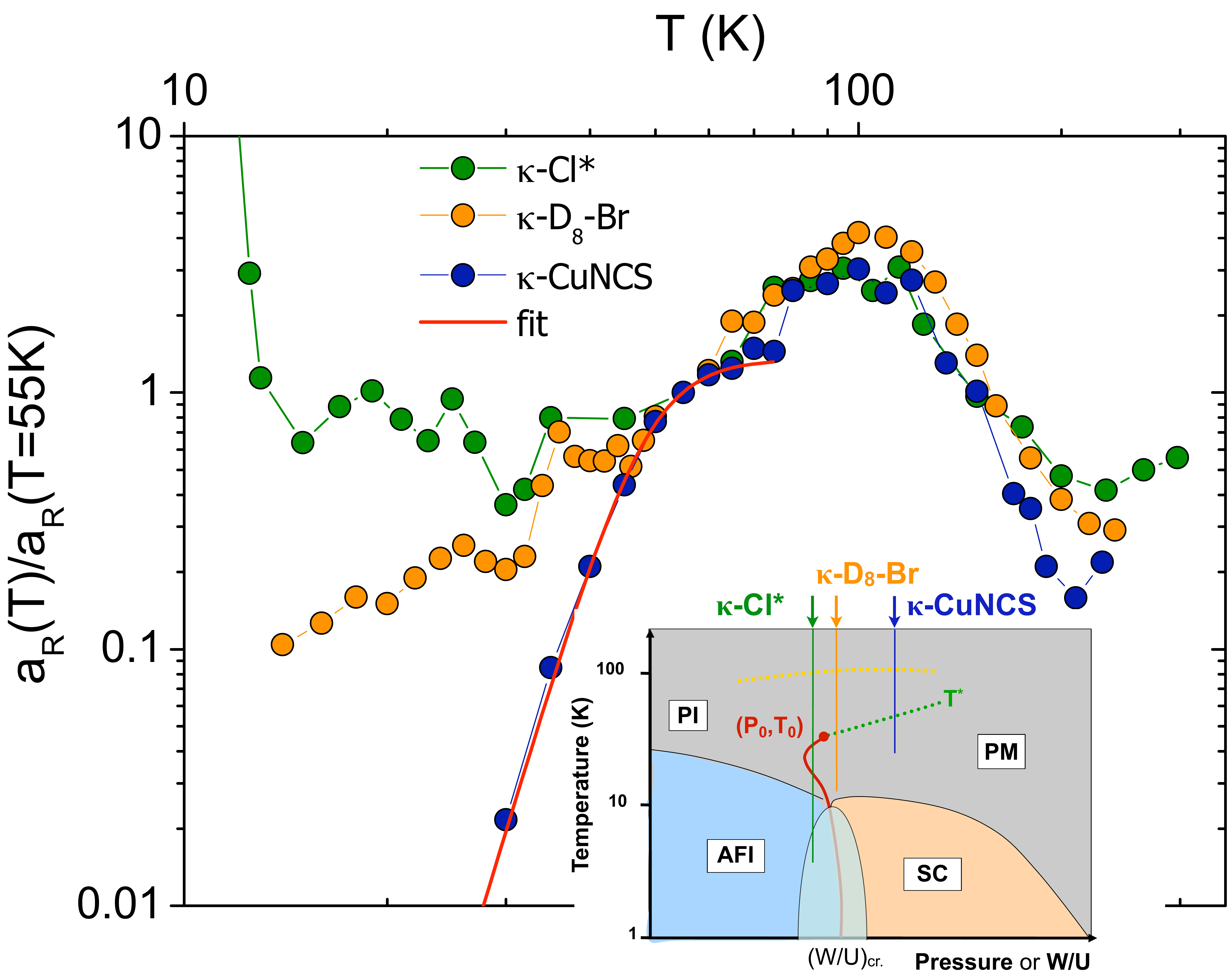}
\caption{\small{(color online) Comparison of the relative noise level \(a_R(T)\) normalized to the value at 55\,K} for three systems located at different positions in the phase diagram, see inset. Dashed yellow line denotes the glass-like ethylene group ordering temperature. \(\kappa\)-Cl\({}^{\ast}\) denotes pressurized \(\kappa\)-Cl.}
\label{fig:abb03}
\end{center}
\end{figure}

A key phenomenon in condensed-matter systems with strong electronic correlations is the Mott metal-to-insulator (MI) transition, i.e.\ the opening of a gap in the charge carrying excitations due to electron-electron interactions \cite{Imada1998}. An important remaining challenge is the understanding of the static and dynamic criticality at the Mott transition, the nature of the anomalous nearby states of electronic matter, and the question of universality \cite{Limelette2003a,Kagawa2005,Kagawa2009}. In particular, superconductivity of unconventional nature and unusual metallic states are frequently observed in the vicinity of the Mott transition, e.g.\ in the high-$T_c$ cuprates and organic charge-transfer salts. Among the latter, the quasi-two-dimensional (2D) $\kappa$-phase salts (BEDT-TTF)$_2$X, where BEDT-TTF (commonly abbreviated as ET) represents bis-ethylenedithio-tetrathiafulvalene and X a polymeric anion, recently have attracted considerable attention as model systems for studying the physics of correlated electrons and the Mott phenomenon in reduced dimensions. 
In these materials, the Mott MI transition and critical endpoint can be easily accessed either by varying the ratio of bandwidth $W$ to effective onsite Coulomb repulsion $U$ by means of physical or chemical pressure or by changing temperature. Accordingly, the ground state of the antiferromagnetic insulator with X = Cu[N(CN)$_2$]Cl$^-$ ($T_N = 27$\,K) can be turned into a superconducting state by either applying a moderate pressure of $\sim 300$\,bar or by modifying the anion: $\kappa$-(ET)$_2$Cu[N(CN)$_2$]Br is a superconductor with $T_c = 11.6$\,K.\\
The generalized phase diagram of these materials \cite{Kanoda1997}, schematically shown in the inset of Fig.~\ref{fig:abb03}, is characterized by a critical region around $(W/U)_{\rm cr.}$, where the S-shaped first-order MI transition line terminates in a second-order critical endpoint, which is reported to be in the temperature-pressure region of (\(p_0 \sim 200 - 300\, \mathrm{bar}, T_0 \sim 30 - 40\, \mathrm{K}\)) \cite{Lefebvre2000,Limelette2003b,Fournier2003,Kagawa2004,Kagawa2005}. In the vicinity of the critical endpoint (\(p_0,T_0\)), unusual metallic behavior characterized by a temperature scale $T^\ast$, where $T_0 \leq T^{\ast}(p > p_0) \simeq 35 - 55\,{\rm K}$, is associated with pronounced anomalies in transport, thermodynamic, magnetic and elastic properties, see~\cite{Toyota2007} and references therein. Among the various explanations that have been proposed for the line of anomalies $T^\ast(p)$ are the formation of a pseudogap in the density of states~\cite{Kataev1992, Mayaffre1994, Kawamoto1995, DeSoto1995}, a crossover from a coherent Fermi liquid at low temperatures into a regime with incoherent excitations (\emph{bad metal}) at high temperatures~\cite{Limelette2003b, Merino2000}, a density-wave instability~\cite{JMueller2002, Sasaki2002, Lang2003}, as well as an incipient divergence of the electronic compressibility caused by the proximity to the Mott transition~\cite{Fournier2003}. Recent thermodynamic investigations suggest that the broadened mean-field like features at \(T^{\ast}\) observed on the metallic side far from the critical point develop into a critical behavior when approaching (\(p_0,T_0\))~\cite{JMueller2002,deSouza2007}. Accordingly \(T^{\ast}\) either marks a line of phase transitions merging into the critical point (\(p_0,T_0\)) of the Mott transition, or \(T^{\ast}(p,X)\) is merely a crossover line, i.e.\ an extension of \(T_{MI}(p,X)\), and the observed effects may be explained in the framework of scaling behavior near the finite-temperature critical endpoint~\cite{Bartosch2010}.

Although the unusual electronic transport properties have been studied in great detail~\cite{Kagawa2005}, there is only limited information about the dynamical properties of the charge carriers, in particular at low frequencies, where interesting effects are to be expected when approaching the critical region of the Mott MI transition.
In this Letter, we report for the first time a correlation-induced, sudden slowing down of the electron dynamics as a fingerprint of the Mott critical point in $\kappa$-(ET)$_2$X, which may be considered as a universal feature of MI transitions. Furthermore, our results indicate a localization-delocalization transition in the interlayer transport when cooling through $T^\ast$ on the metallic side of the phase diagram, contributing to the yet unsettled debate about the nature of the phase above $T^\ast$.

Single crystals of \(\kappa\)-(ET)\({}_{2}\)X with \(\mathrm{X = Cu[N(CN)_{2}]Cl^-}\), Cu[N(CN)\({}_{2}\)]Br$^-$ and Cu(NCS)\({}_{2}\)$^-$ (hereafter $\kappa$-Cl, $\kappa$-Br and $\kappa$-CuNCS, respectively), as well as the fully deuterated variant \(\kappa\)-(D\({}_{8}\)-ET)\({}_{2}\)Cu[N(CN)\({}_{2}\)]Br (\(\kappa\)-D\({}_{8}\)-Br) of plate- and rod-shaped morphology were grown by electrochemical crystallization~\cite{Williams1990,Wang1990,Urayama1988}. Low-frequency fluctuation spectroscopy measurements have been performed in a five-terminal setup using a standard bridge-circuit ac technique. The experiment is described in detail in~\cite{Mueller2011}. Care has been taken that spurious noise sources, in particular contact noise, do not contribute to the results. Resistance ($R$) and noise power spectral density ($S_R$) have been measured perpendicular to the conducting layers.

\begin{figure}[htbp]
\begin{center}
\includegraphics[width=0.425\textwidth]{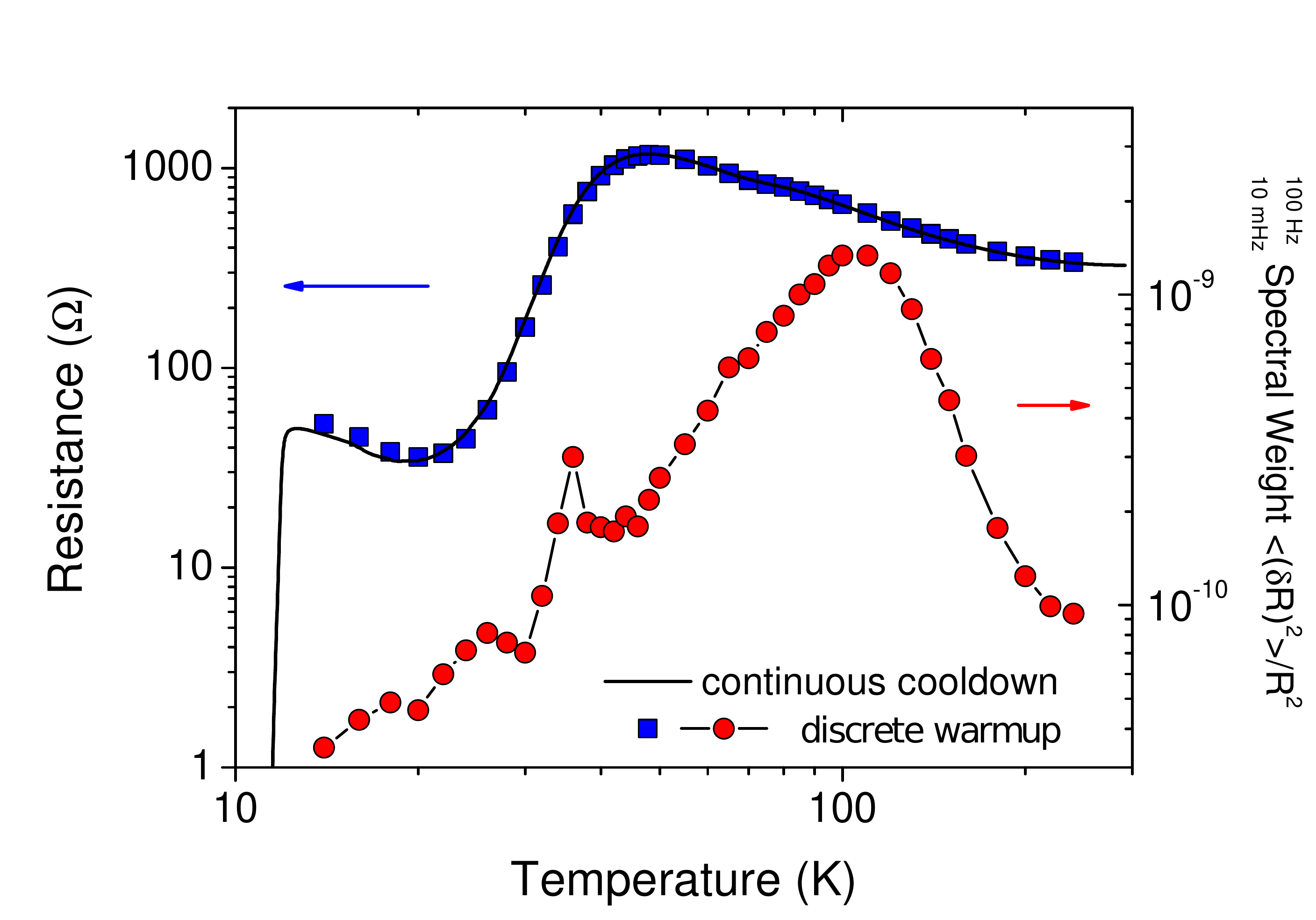}
\caption{\small{(color online) Resistance and integrated resistance noise power of \(\kappa\)-D\({}_{8}\)-Br.} Noise spectra have been taken in discrete steps upon warming the sample after slow continuous cooldown.}
\label{fig:abb01}
\end{center}
\end{figure}
For all samples, at the temperatures of interest here, we have observed excess noise of general $1/f^\alpha$-type, characterizing the intrinsic resistance (conductance) fluctuations. Representative for our experiments, Fig.~\ref{fig:abb01} shows the resistance \(R\) and the normalized spectral weight of the resistance fluctuations, $\langle (\delta R)^2\rangle/R^2 = \int S_R(f)/R^2 {\rm d}f$, integrated over a bandwidth $0.01 - 100$\,Hz for a deuterated sample \(\kappa\)-D\({}_{8}\)-Br. The investigated samples are continuously cooled down to 4.2\,K with a slow cooling rate of 3\,K/h. Noise spectra have been taken while warming up the sample in discrete steps. For control, the resistance values are determined simultaneously and are found to be in very good agreement with the continuous resistance measurement (Fig.~\ref{fig:abb01}, solid line). 
The \(\kappa\)-D\({}_{8}\)-Br sample investigated here, is situated very close to the Mott transition on the metallic side of the phase diagram (Fig.~\ref{fig:abb03}, inset). At high temperatures, the resistance shows a semiconducting behavior down to about 50\,K, below which a step-like decrease of the resistance by almost two orders of magnitude indicates the occurrence of a metallic phase. Below 20\,K, the resistance shows an increase again (with small hysteresis), which may be caused by electron localization near the Mott transition~\cite{Sano2010}, before the transition into the superconducting state occurs at a temperature \(T_c = 12.6\)\,K (onset).
As becomes clear from Fig.~\ref{fig:abb01}, the temperature dependence of the spectral weight of the noise power looks remarkably different than \(R(T)\). Below a broad Gaussian-shaped maximum centered around \(T \simeq 100\)\,K, $S_R(T)$ slowly decrease with decreasing temperature. Superimposed on this general trend is a pronounced anomaly at around $35 \sim 36$\,K, a temperature that is very close to the critical endpoint (\(p_0,T_0\)) of the Mott transition, followed by two smaller peaks at lower temperatures. Besides the noise power spectral density \(S_R\) (magnitude of the fluctuations), the frequency exponent of the \(1/f^{\alpha}\)-type noise, $\alpha(T) = - \partial \ln{S_R(f,T)}/\partial \ln{f}$, can be extracted from the measured spectra.  
\begin{figure}[htbp]
\begin{center}
\includegraphics[width=0.35\textwidth]{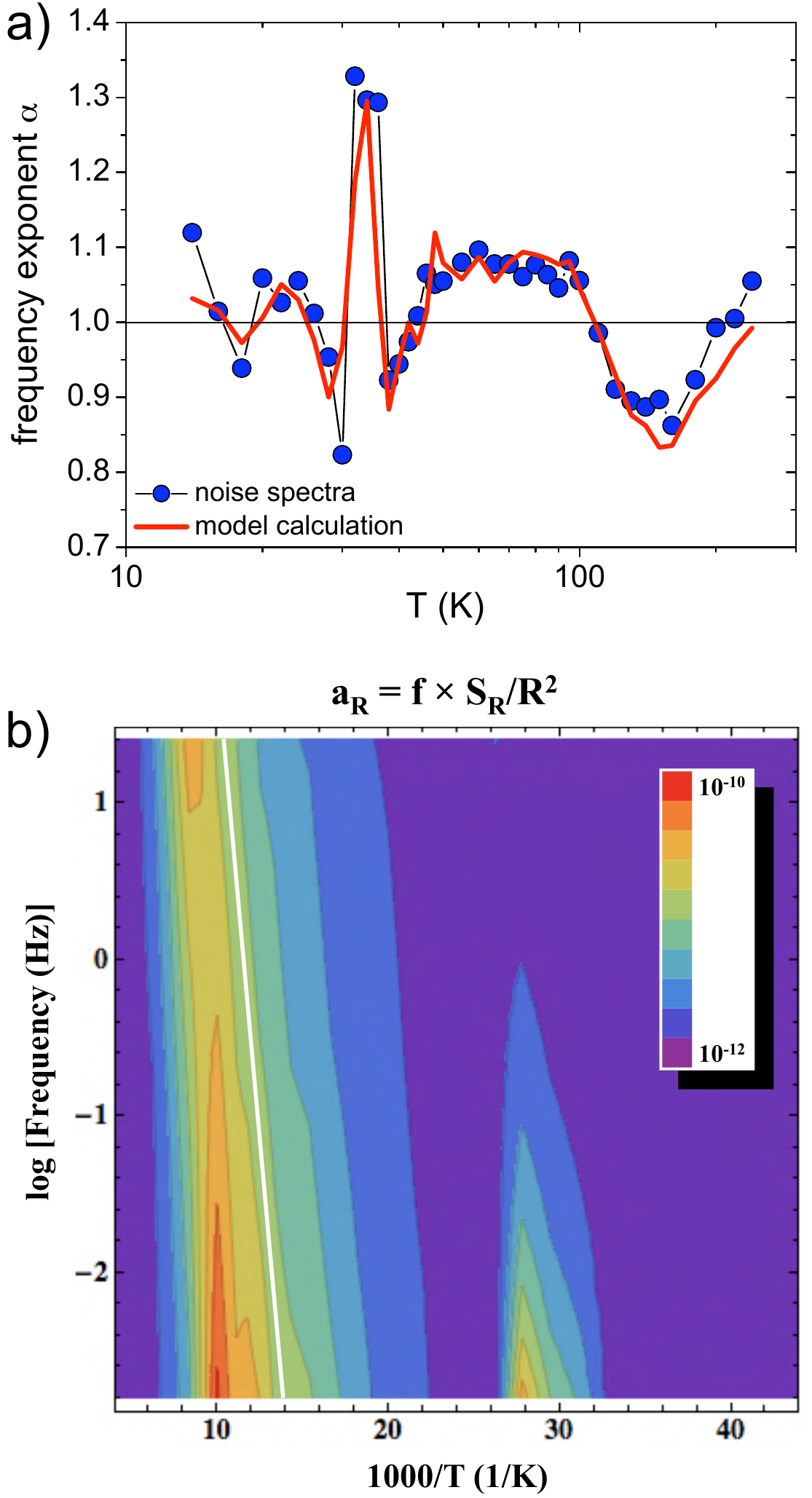}
\caption{\small{(color online) a) Measured frequency exponents, \(\alpha_{\rm meas.}(T)\) (symbols) compared to the predicted values \(\alpha_{\rm calc.}(T)\) (line), for \(\kappa\)-D\({}_{8}\)-Br in the framework of the generalized DDH-model \cite{Raquet1999}. b) Contour plot of the relative noise level $a_R$ vs.\ frequency $f$ and temperature $T$ in an Arrhenius representation. The white line corresponds to the slope $E_a = 250$\,meV as derived from the DDH model.}}
\label{fig:abb02}
\end{center}
\end{figure}
We find a strongly non-monotonic behavior in $\alpha(T)$, shown in Fig.~\ref{fig:abb02}~a), indicating substantial variations of the electrons' dynamical properties in different temperature regimes. 
Our previous studies have revealed that the temperature dependence of the nearly \(1/f\)-type noise spectra can be well described by a generalized random fluctuation model, see~\cite{Raquet1999,JMueller2009b,Mueller2011}, first used by Dutta, Dimon and Horn (DDH) to describe noise processes in metals~\cite{Dutta1979}. The phenomenological model assumes thermally-activated fluctuators --- not specified \emph{a priori} --- which linearly couple to the resistance. The \(1/f^{\alpha}\)-type noise is produced by a large number of fluctuators acting in concert (each contributing a Lorentzian spectrum), whereas the distribution of activation energies \(D(E)\) governs both the strong temperature dependence of the magnitude of the noise and any deviations from perfect (i.e.\ \(\alpha = 1\)) \(1/f\) behavior. In this case, unlike for an individual fluctuator, the kinetics is nonexponential, i.e.\ the correlation function falls off with time not exponentially~\cite{Kogan1996}, and:
\begin{equation}
\frac{S_R(f)}{R^2} = \frac{1}{\pi f} \int^{\infty}_{0} \mathrm{d}E \frac{D(E,T)}{\cosh[(E-E_{\omega})/k_B T]}.
\label{eq1}
\end{equation}
Due to the large logarithmic factor in \(E_{\omega} = -k_B T \ln{\omega \tau_0}\) (\(\tau_0\) is an attempt time in the order of inverse phonon frequencies, typically $10^{-12} - 10^{-14}$\,s), ordinary activation energies in solids can be accessed with this technique~\cite{Kogan1996}. If \(D(E)\) is not an explicit function of temperature, the assumptions of the DDH model can be checked for consistency by comparing the measured frequency exponent \(\alpha_{\rm meas}(T)\) with the predictions of the model \cite{Dutta1979}:
\begin{equation}
\alpha_{\rm calc}(T) = 1 - \frac{1}{\ln \omega \tau_0} \biggl(\frac{\partial \ln S(f,T)}{\partial \ln T} - 1\biggr).
\label{nonexp}
\end{equation}
As shown in Fig.~\ref{fig:abb02}a), the agreement of the measured data with the generalized DDH model is  . In particular, the model calculation provides a good quantitative description of the measured $\alpha(T)$ and accounts for all of the features, i.e.\ non-monotonic temperature dependences being related to the fingerprints of the fluctuating entities, since for $\alpha$ greater or smaller than 1, $\partial D(E)/\partial E > 0$ and $\partial D(E)/\partial E < 0$, respectively. 
The striking agreement of the measured data with the model calculation underlines the physical meaning of extracting \(D(E)\) from the temperature dependence of the magnitude of the noise by $D(E) \propto S_R/R^2 \times 2\pi f/k_B T$. From this analysis (not shown) we find that \(D(E)\) exhibits a pronounced maximum at about $E_\omega \sim 250$\,meV, which corresponds to the activation energy of the orientational degrees of freedom of the ET molecules' terminal ethylene groups, as determined e.g.\ by NMR, specific heat and thermal expansion~\cite{Miyagawa1995, Akutsu2000, JMueller2002}, i.e.\ in this case the relevant fluctuators can be identified \emph{a posteriori}  \cite{JMueller2009b}.\\
The noise results on $\kappa$-D$_8$-Br are summarized in a contour plot of the relative noise level $a_R(f,T) = f \times S_R/R^2$ shown in Fig.~\ref{fig:abb02}b) in an Arrhenius plot. The enhanced noise level at elevated temperatures (around 100\,K at $f = 1$\,Hz) seen in the entire accessible frequency range are due to the coupling of the electronic fluctuations to the abovementioned conformational motion of the ET molecules' ethylene endgroups. The main result, however, is the abrupt and strong enhancement of $a_R$ at $T^\ast \sim 36$\,K occuring at the lowest measuring frequencies which we interpret as a sudden slowing down of the electron dynamics caused by the vicinity to the critical endpoint at $T_0 \approx 35 \sim 36$\,K.

A systematic investigation of different compounds reveals evidence that the enhanced low-frequency fluctuations for $\kappa$-D$_8$-Br are induced by electronic correlation effects: Fig.~\ref{fig:abb03} shows the relative noise level \(a_R = \overline{f} \times S_R / R^2\) at a bandwidth \(\overline{f} = 1\)\,Hz for three different compounds: pressurized $\kappa$-Cl (denoted as $\kappa$-Cl$^\ast$, similar sample as in \cite{JMueller2009a}) and $\kappa$-D$_8$-Br, both of which are located close to the Mott critical region $(W/U)_{\rm cr.}$, and $\kappa$-CuNCS on the far metallic side of the phase diagram. Metallic $\kappa$-H$_8$-Br (not shown) behaves similarly than the latter system. In order to better compare the different systems, we show \(a_R(T)\) normalized to the value at 55\,K. Strikingly, all three systems show a similar behavior above about 50\,K. This indicates that the same processes are causing the excess noise for $T >  50$\,K, independent of the materials' position in the phase diagram. A common feature is the strong coupling of the charge carriers' fluctuation properties to the conformational modes of the ET molecules' ethylene endgroups. In contrast, as seen in Fig.~\ref{fig:abb03}, for \(T < 50\)\,K the temperature characteristics of \(a_R\) strongly depends on the position in the phase diagram. For the metallic system \(\kappa\)-CuNCS, a strong (exponential) decrease of the noise level of two orders of magnitude is observed, which can be well fitted by a step function (a logistic function is shown), indicating a transition or crossover of two qualitatively different transport regimes. Interestingly, the midpoint of this crossover at 49\,K determined from the fit to the data corresponds to the $T^\ast$ anomaly in thermodynamic measurements~\cite{Lang2003}. 
The vanishing $1/f$ noise below $T^\ast$ is consistent with 3D coherent electronic transport dominated by the conducting properties of the ET layers, whereas above $T^\ast$ interlayer hopping processes play a major role. In the transition region, the conducting sheets become increasingly decoupled for increasing temperature, which leads to the observed step-like increase in \(a_R(T)\). Hopping as well as excitations over a band gap have been discussed in early models of interlayer transport at elevated temperatures~\cite{Cariss1990, Toyota1990}. Activation kinetic processes of this kind are a prerequisite for the concept of nonexponential kinetics, which may explain the excellent correspondence between the experimental noise data and the DDH model as well as the observation that \(a_R(T > 50\,{\rm K})\) is independent of the position in the phase diagram. 
Such an coherent-incoherent electronic transport transition/crossover is in very good agreement with recent experiments of the conduction electron spin resonance (CESR), where for \kCl\ and \kBr\ it is found that at high temperatures perpendicular transport is strongly incoherent and that diffusion is confined to single molecular layers~\cite{Antal2009}. According to these measurements, the electrons diffuse several hundreds of nm without interlayer hopping within the spin lifetime of 10\({}^{-9}\)\,s. Below 50\,K, an "unexplained line broadening" \cite{Antal2009} may indicate the {\em bad metal} to normal metal, i.e.\ a 2D to 3D spin diffusion, crossover in agreement with our results.

In contrast to the metallic systems, the noise of the more correlated \(\kappa\)-D\({}_{8}\)-Br shows a pronounced anomaly at around $35 \sim 36$\,K and exhibits a roughly linear temperature dependence for \(T < 30\)\,K, superimposed to which are smaller peak-like features. The relation of those to the recently discussed pseudogap state \cite{Shimizu2010} is still a matter of investigation. A very similar behavior is observed in \(\kappa\)-Cl\({}^{\ast}\), but there -- due to the slightly different (more left) position in the phase diagram -- for \(T < 30\)\,K the normalized resistance noise power \(a_R(T)\) remains on a larger and constant level, apart from the strong increase near \(T_c\), which is due to the percolative nature of the superconducting transition~\cite{JMueller2009a}. Considering the position in the phase diagram, i.e.\ the strength of electronic correlations as being the major difference between the systems shown in Fig.~\ref{fig:abb03}, the increased noise level for \(\kappa\)-D\({}_{8}\)-Br below about 50\,K has to be attributed to the vicinity of the critical endpoint at $(p_0 = (W/U)_{\rm cr.}, T_0 \approx 35 \sim 36\,{\rm K})$.  Importantly, Figs.~\ref{fig:abb02} and \ref{fig:abb03} reveal that the correlation-induced, strongly enhanced noise level close to $(p_0,T_0)$ goes along with a substantial shift of spectral weight to lower frequencies, i.e.\ a slowing down of the electron dynamics when approaching the critical endpoint both from below and above. Similar observations, i.e.\ strongly enhanced low-frequency fluctuations have been made for the critical region of other canonical MI transitions, namely the Anderson-Mott transition in P-doped Si \cite{Kar2003}, as well as carrier-concentration and magnetic-field-induced MI transition in Si inversion layers \cite{Bogdanovic2002,Jaro2002}. Our observations offer the intriguing possibiliy to explore the dynamical scaling properties and possible correlated electron dynamics as suggested for other MI transitionss \cite{Kar2003} directly at the critical point in pressure-tuned \kCl.

\begin{acknowledgments}
Work supported by the Deutsche Forschungsgemeinschaft (DFG) through the Emmy Noether program and SFB/TR 49. Work at Argonne National Laboratory is supported by the U. S. Department of Energy Office of Science under Contract No. DE-AC02-06CH11357.
\end{acknowledgments}

\bibliographystyle{unsrtnat}

\end{document}